\documentclass[twocolumn,showpacs,aps,prl,superscriptaddress,floatfix,letterpaper]{revtex4}

\usepackage{graphicx}
\usepackage{dcolumn}
\usepackage{amsmath}
\usepackage{epsfig}

\input babarsym

\newcommand{\BaBarType}      {PUB}  % Journal publication
\newcommand{\BABARPubYear}    {08}
\newcommand{\BABARPubNumber}  {018}
\newcommand{\SLACPubNumber} {13230}

\def\figurebox#1#2#3{%
    \def\arg{#3}%
    \ifx\arg\empty
    {\hfill\vbox{\hsize#2\hrule\hbox to #2{\vrule\hfill\vbox to #1{\hsize#2\vfill}\vrule}\hrule}\hfill}%
    \else
    {\hfill\epsfbox{#3}\hfill}%
    \fi}

\def\Resultrhoprhom{8.3\pm 0.7 (\mathrm{stat})\pm 0.8(\mathrm{syst}) ~\mathrm{fb}}
\def\ResultrhoprhomExtend{19.5\pm 1.6 (\mathrm{stat})\pm 3.2(\mathrm{syst})~\mathrm{fb}}
\def\Myamplitudes{|{F}_{00}|^2:|{F}_{10}|^2:|{F}_{11}|^2=0.51\pm0.14(\mathrm{stat})\pm0.07(\mathrm{syst}):0.10\pm0.04 (\mathrm{s
tat})\pm0.01(\mathrm{syst}):0.04\pm0.03(\mathrm{stat})\pm0.01(\mathrm{syst}) }

\begin{document}

\begin{flushleft}
\ \hfill\ \\
\ \hfill\ \\
\ \hfill\ \\
\hfill \babar\/-\BaBarType-\BABARPubYear/\BABARPubNumber \\
\hfill SLAC-PUB-\SLACPubNumber 
\end{flushleft}

\title
{
{\large \bf
\boldmath{Observation of $e^+e^-\rightarrow$  $\rho^+\rho^-$ near $\sqrt{s}=10.58$\gev}
} 	
}
%% author list as of 05-May-2008 (530 authors)
%
\author{B.~Aubert}
\author{M.~Bona}
\author{Y.~Karyotakis}
\author{J.~P.~Lees}
\author{V.~Poireau}
\author{E.~Prencipe}
\author{X.~Prudent}
\author{V.~Tisserand}
\affiliation{Laboratoire de Physique des Particules, IN2P3/CNRS et Universit\'e de Savoie, F-74941 Annecy-Le-Vieux, France }
\author{J.~Garra~Tico}
\author{E.~Grauges}
\affiliation{Universitat de Barcelona, Facultat de Fisica, Departament ECM, E-08028 Barcelona, Spain }
% ...
\collaboration{The \babar\ Collaboration}
\noaffiliation

\date{\today}% It is always \today, today, but you may specify any date with \date.

\begin{abstract}

We report the first observation of \epem\to$\rho^+\rho^-$,
in a data sample of 379~\invfb  collected with the \babar\/\ detector at
the  \pep2 $e^+e^-$ storage ring at center-of-mass energies near
$\sqrt{s}\! =$10.58~\gev.
We measure a cross section
of $\sigma(\epem\!\! \to\! \rho^+ \rho^-)\! =$$\ResultrhoprhomExtend$.
Assuming production through single-photon annihilation, there are three independent helicity amplitudes.
We measure the ratios of their squared moduli to be $\Myamplitudes$.
The $|{F}_{00}|^2$ result is inconsistent with the prediction of 1.0 
made by QCD models with a significance of 3.1 standard deviations including 
systematic uncertainties.

\end{abstract}

\pacs{13.66.Bc, 13.25.-k, 14.40.Ev}

\maketitle

The exclusive production of $J/\psi \eta_c$ and other double-charmonium vector-pseudoscalar (VP) pairs 
in \epem collisions around the $\Upsilon(4S)$ mass ($\sqrt s \approx 10.58 \gev$)   
is observed~\cite{belledoublecc,babardoublecc}
at rates approximately ten times larger than the rates expected from QCD-based
models~\cite{nrqcd}. Various theoretical efforts have been made to resolve the
discrepancy~\cite{aclightcone}.
Measurements of the process \epem\to$\phi\eta$~\cite{babarphieta}
provide information on the \epem\to  VP process in the strange quark sector.
Study of the vector-vector (VV) process \epem\to$\rho^+\rho^-$ can provide
complementary information and
test perturbative QCD at the amplitude level~\cite{stanrule} 
through investigation of the VV angular distributions.

The charge-conjugation ($C$) even final states $\rho^0\rho^0$ and $\phi\rho^0$ are
produced through \epem  two--virtual-photon annihilation (TVPA)~\cite{babarvv,davier,bodwinvv}.
For $\rho^+\rho^-$, $C$ can be either positive or negative. However,
due to the particles' charges, the \epem\to $\rho^+\rho^-$ process is unlikely to occur via TVPA
unless there is either significant final quark recombination between the products of the two 
virtual photons or final-state interactions (FSI)~\cite{reggeo}. 
Assuming production through single-photon annihilation or $\Upsilon(4S)$ decay, the VV final state 
can be described with three independent helicity amplitudes. Any discrepancy between
the amplitudes predicted by perturbative QCD and the experimental measurement might indicate contributions from mechanisms such as FSI. Such discrepancies could help to better understand the 
importance of FSI effects in $B\to VV$ decays~\cite{bfsi}.

This analysis uses  343~\invfb  of \epem  data collected
on the \Y4S resonance at 10.58~\gev and 36\invfb collected 40~\mev below (off-resonance)
with the \babar\/\ detector at the SLAC PEP-II asymmetric-energy $B$ factory.
The \babar\/\ detector is described in detail elsewhere~\cite{babardetector}.
Charged-particle momenta and energy loss are measured in the
tracking system, which consists of a silicon vertex tracker (SVT) and a helium-isobutane drift chamber (DCH).
Electrons and photons  are detected in a CsI(Tl) calorimeter (EMC).
Charged pion candidates are identified using likelihoods of 
specific ionization in the
SVT and DCH, and of Cherenkov angle and photon counts measured in an 
internally reflecting ring-imaging Cherenkov detector.
Photons are identified by clusters of energy deposited in the EMC
that have shapes consistent with an electromagnetic shower. The clusters
are required to be isolated, i.e., geometrically unassociated with charged tracks.

To form the $\rho^+\rho^-$ final state, we select
events with exactly two well-reconstructed oppositely-charged $\pi^{\pm}$ and at least
two well-reconstructed $\pi^0$ candidates.
We require the $\pi^{\pm}$ candidates to have at least 12 DCH hits and
a laboratory polar angle well within the SVT acceptance of $0.41<\theta<2.54$
radians. The laboratory transverse momenta of the $\pi^{\pm}$ 
candidates are required to be greater than $100~\mevc$.
The two charged tracks must both be identified as pions.
We fit the two charged tracks to a common vertex, and require the $\chi^2$
probability to exceed 0.1\%.

The photon candidates used to reconstruct $\pi^0$ candidates
are  required to have a minimum laboratory energy of 100 MeV.
The invariant masses of the candidate photon pairs are required to be within [0.1, 0.16] \gevcc. 
The masses of $\pi^0$ candidates are then constrained to the world average value~\cite{pdg06}.

The $\rho^{\pm}$ candidates are formed by combining a $\pi^{\pm}$ candidate with a $\pi^0$ candidate.
The production angle $\theta^*$ is defined as the angle 
between the $\rho^+$ meson direction and the
incident $e^-$ beam in the \epem center-of-mass.  
The $\rho^{\pm}$ helicity angles $\theta_{\pm}$
are defined as the angles in the $\rho^{\pm}$ rest frame between the
direction of the boost from the laboratory frame and the direction of the $\pi^{\pm}$.
We require $|\cos\theta^*|<0.8$ and $|\cos\theta_{\pm}|\! <\! 0.85$ 
because there is  low signal efficiency outside this fiducial region.

\begin{figure}
\begin{center}
\includegraphics[width=7.0cm]{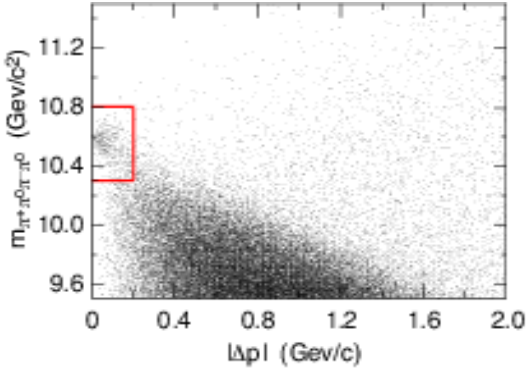}
\caption{
Scatter plot of $m_{\pip\pi^0\pim\pi^0}$ {\it vs.} $|\Delta p|$
between the $\pip\pi^0\pim\pi^0$  and initial \epem systems for the on-resonance data.
% (Color online)
}
\label{fig:4trkmassscatter}
\end{center}
\end{figure}

Figure~\ref{fig:4trkmassscatter} shows
the scatter plot of the invariant mass $m_{\pi^+\pi^0\pi^-\pi^0}$ versus the absolute momentum 
difference $|\Delta p|$ in the laboratory frame between the $\pi^+\pi^0\pi^-\pi^0$  
and  initial \epem systems after requiring the $\pi^{\pm}\pi^0$ masses to be less than 1.5 \gevcc.
The last requirement eliminates a twofold ambiguity in forming the $\rho^{\pm}$ candidates.
A few per cent of the events have more than one $\rho^+$ or $\rho^-$ 
candidate because of multiple $\pi^0$'s.
All candidates are kept.

\begin{figure}
\begin{center}
\includegraphics[width=7.0cm]{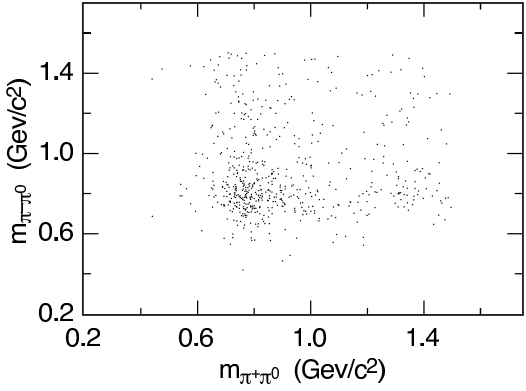}
\caption{
Scatter plot of $m_{\pip\pi^0}$ and $m_{\pim\pi^0}$ for the accepted events in the combined data.% (Color online)
}
\label{fig:massmassscatter}
\end{center}
\end{figure}

We accept events from within the rectangular area indicated in Fig.~\ref{fig:4trkmassscatter} 
($|m_{\pi^+\pi^0\pi^-\pi^0}-\sqrt{s}|<0.28$ \gevcc and $|\Delta p|<0.2$ \gevc).
There are a total of $612$ candidates from $571$ events in the $\Y4S$ and off-resonance samples combined.
Figure~\ref{fig:massmassscatter} shows the scatter plot of the invariant masses of $\pi^+\pi^0$ and $\pi^-\pi^0$
pairs from the accepted candidates.
The concentration of candidates in the $\rho^+\rho^-$ mass range indicates $\rho^+\rho^-$ production.

We use a two-dimensional maximum likelihood fit to extract the signal for \epem\to$\rho^+\rho^-$.
Since the final state particle masses are far below
the \epem collision energy, we treat the two-body masses as uncorrelated.  The signal
probability density function (PDF) is constructed as a product of two identical one-dimensional
PDFs for $\rho^{+}$ and $\rho^{-}$.
We use a $P$-wave relativistic Breit-Wigner formula to construct the PDF for the $\rho^{\pm}$ resonance:

\begin{equation}
F(m) \propto \frac{m\Gamma(m)}{({m_0}^2-m^2)^2+{m_0}^2\Gamma^2(m)} ,
\end{equation}

$$
\Gamma(m) = \Gamma_0\left(\frac{q}{q_0}\right)^3\left(\frac{m_0}{m}\right)\left(\frac{1+ q_0^2 R^2}{1+ q^2 R^2}\right) ,
$$
where $m$ is the observed pion-pair mass,
$\Gamma$ is the mass-dependent $\rho$ width, 
and $q$ is the absolute value of the pion candidate momentum in the $\rho$ candidate rest frame.
The $0$ subscript indicates the value at the central mass of the $\rho$ resonance.
$R$ is the Blatt-Weisskopf damping radius, which we set to 3 $(\gevc)^{-1}$\cite{aston,godfrey}.

A threshold function $q^3/(1+q^3\alpha)$ is used to model the background in the $\rho^{\mp}\pi^{\pm}\pi^0$
system, where $\alpha$ is a shape parameter.
We use a linear function to model the residual two-dimensional background:
\begin{equation}
B(m_{\pi^+\pi^0},m_{\pi^-\pi^0})=1+ a(m_{\pi^+\pi^0}-M)+a(m_{\pi^-\pi^0}-M) ,
\end{equation}
where  $a$  is a floating parameter, and $M = 0.89 \gevcc$ is the midpoint
of the $\pi^\pm\pi^0$ invariant mass range used in the fit.

\begin{figure*}
\begin{center}
\includegraphics[width=14.0cm]{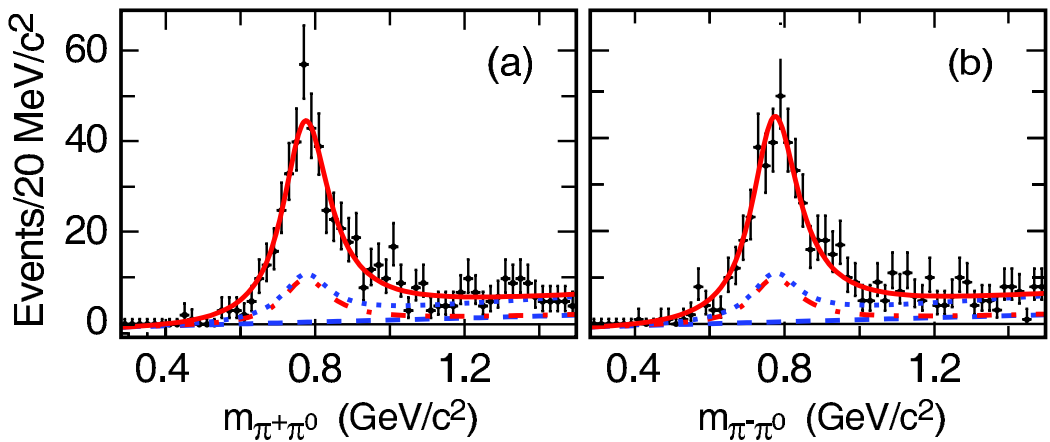}
\caption{
The invariant mass projections a) $m_{\pi^+\pi^0}$ ($m_{\pi^-\pi^0}<1.5$ \gevcc) and
b) $m_{\pi^-\pi^0}$ ($m_{\pi^+\pi^0}<1.5$ \gevcc) for accepted events in the combined data.
The blue-dashed line is the residual linear background, the red dot-dashed line adds
a) $\rho^{+} \pi^{-}\pi^0$,  b) $\rho^{-} \pi^{+}\pi^0$, and the blue-dotted line includes 
both $\rho^{+} \pi^{-}\pi^0$  and $\rho^{-} \pi^{+}\pi^0$.
The red solid line adds the signal.  % (Color online)
 }
\label{fig:massprojection}
\end{center}
\end{figure*}

In the fit to the data, we fix the  mass and width of the $\rho^{\pm}$
to the world average values~\cite{pdg06}.
The parameters varied in the fit are: 
$\alpha$ [$\alpha(\pi^+\pi^0)$=$\alpha(\pi^-\pi^0)$], $a$, and the
numbers of events for the four components: $\rho^+\rho^-$, $\rho^+\pi^-\pi^0$, $\rho^-\pi^+\pi^0$, 
and the residual background.
The mass projections on $m_{\pi^+\pi^0}$ and $m_{\pi^-\pi^0}$ from
the two-dimensional fit are shown in Fig.~\ref{fig:massprojection}.
The extracted  number of $\rho^+\rho^-$ signal events  is $357\pm 29$,
with $329\pm 25$ in the $\Upsilon(4S)$ resonance sample and $31\pm 14$ in the off-resonance sample.

Assuming that the $\rho^+\rho^-$ is  produced through a  $J^{PC}=1^{--}$ object
(a single-photon or $\Upsilon(4S)$), and that C and parity P are conserved, 
there are three independent complex helicity amplitudes, ${F}_{00}$, ${F}_{10}$, and ${F}_{11}$, 
where the indices indicate the helicities of the $\rho$ mesons.
${F}_{10}={F}_{\pm10}={F}_{0\pm1},~{F}_{11}={F}_{-1-1}$,
and ${F}_{1-1}={F}_{-11}=0$ due to angular momentum conservation~\cite{chung}.
The angular distribution of $\rho^+\rho^-$ decay products can be expressed as:
$$
\frac{dN}{d \cos\theta^* d\cos\theta_{+} d\varphi_{+} d\cos\theta_{-} d\varphi_{-}} \propto  |A_{+1}|^2+|A_{-1}|^2 , \\
$$
\begin{eqnarray}
A_{\pm 1} & = & \sin\theta^*\cos\theta_+\cos\theta_- |F_{00}| \nonumber \\
 & & +\sin\theta^*\sin\theta_+\sin\theta_-\cos(\varphi_+ + \varphi_-) |F_{11}|e^{i\varphi_{11}} \nonumber \\
 & & +\frac{1}{2}\sin\theta_+\cos\theta_-(\pm(1\mp\cos\theta^*)e^{i\varphi_+} \nonumber  \\
 & &                                 \mp(1\pm\cos\theta^*)e^{-i\varphi_+})|F_{10}|e^{i\varphi_{10}} \nonumber  \\ 
 & & +\frac{1}{2}\cos\theta_+\sin\theta_-(\pm(1\mp\cos\theta^*)e^{i\varphi_-}  \nonumber \\
 & &                                  \mp(1\pm\cos\theta^*)e^{-i\varphi_-})|F_{10}|e^{i\varphi_{10}} ,
\label{phietaang}
\end{eqnarray}
where $\varphi_{\pm}$  is the azimuthal angle that corresponds to the helicity (polar) angle $\theta_{\pm}$
defined above. 
In this coordinate system, the incoming electron direction has an azimuthal angle $\varphi_{\pm}$ of zero.
The angles $\varphi_{11}$ and $\varphi_{10}$ are the strong phases of the amplitudes.
Due to limited statistics, we examine only the projections and thus lose sensitivity to these phases.

The one-dimensional angular distributions are obtained from Eq.~(\ref{phietaang}) by integrating over all other angles.
When integrating over the full angular range, the results are
\begin{eqnarray}
\lefteqn{\frac{dN}{d \cos\theta^* } \propto \sin^2\theta^* |F_{00}|^2 } & \nonumber \\
         &  +f_1 (1+\cos^2\theta^*)|F_{10}|^2 +f_2  \sin^2\theta^*|F_{11}|^2\ \ 
\label{phietaprodang},
\end{eqnarray} 
\begin{eqnarray}
\lefteqn{\frac{dN}{d\cos\theta_{\pm}} \propto \cos^2\theta_{\pm}|F_{00}|^2 } & \nonumber \\
                    &  + (f_3+f_4 \cos^2\theta_{\pm})|F_{10}|^2+f_5 \sin^2\theta_{\pm}|F_{11}|^2\ \ 
\label{phietaphihelicityang},
\end{eqnarray}
\begin{equation}
\frac{dN}{ d\varphi_{\pm}} \propto |F_{00}|^2 + (f_6-f_7 \cos 2 \varphi_{\pm} )|F_{10}|^2+ f_2 |F_{11}|^2 ,
\label{phietavarphiang}
\end{equation}
where the constants $f_n$ are given in the first row of Table~\ref{tab:angconst}.
\begin{table}[!htb]
\caption{Constants in equations~\ref{phietaprodang},~\ref{phietaphihelicityang}, and~\ref{phietavarphiang}.}
\begin{center}
\begin{tabular}{lrrrrrrr}
\hline
Integrated region  & $f_1$ & $f_2$ & $f_3$ & $f_4$ & $f_5$ & $f_6$ & $f_7$ \\
\hline\hline
Full range         &  2    & 2     & 1     & 1     & 1     & 4     & 1     \\
Fiducial region    &  3.15 & 4.97  & 0.77  & 1.66  & 1.58  & 6.44  & 3.15  \\
\hline\hline
\end{tabular}
\label{tab:angconst}
\end{center}
\end{table}

To determine the amplitude factors, we perform fits of
Eqs.~\ref{phietaprodang},~\ref{phietaphihelicityang}, and~\ref{phietavarphiang}
to the data.  The fits are performed in the fiducial region $|\cos\theta^*|<$ 0.8 and $|\cos\theta_{\pm}|\! <\! 0.85$.
Limiting the integration to this fiducial region yields the constants shown in the second row of the table.

We use the sPlot~\cite{sweight} technique to subtract backgrounds in the measured angular distributions.
This technique assigns a weight to each event (sWeight) for each category to which it might belong.
The sWeights are obtained from the 2D fit to the $m_{\pi^-\pi^0}$ versus $m_{\pi^+\pi^0}$ distribution.
We subdivide $\cos\theta^*$ and $cos\theta_\pm$ into bins and produce an efficiency table from a
phase-space-based Monte Carlo (MC) simulation. The event weight is given by
the sWeight divided by the efficiency. 

\begin{figure}
\begin{center}
\includegraphics[width=8.5cm]{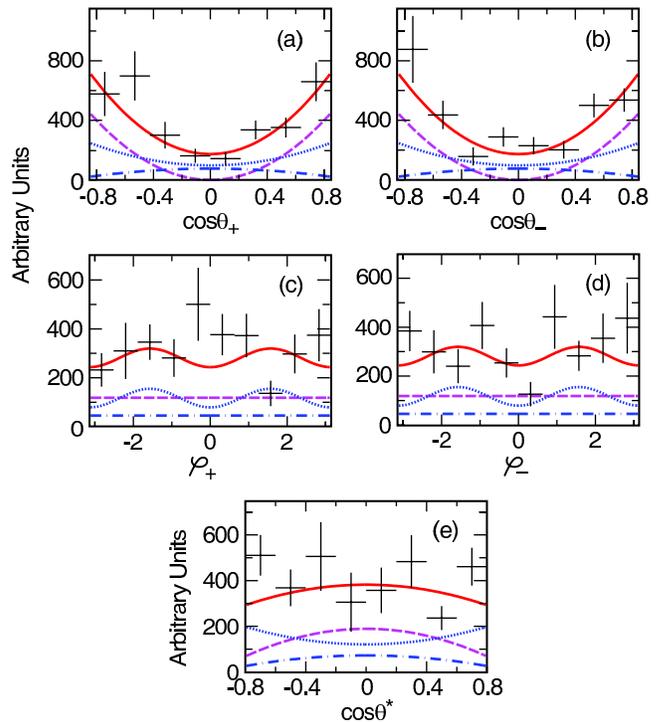}
\caption{ 
The background-subtracted (sWeighted) and efficiency-corrected  a) $\cos\theta _+$ b) $\cos\theta _-$
c) $\varphi_+$  d) $\varphi_-$  and e) $\cos\theta^*$
distributions.  The magenta dashed line is the contribution from the ${F}_{00}$ component, the
blue dotted line is for the ${F}_{10}$ component, the blue dot-dashed line is for the ${F}_{11}$ component,
and the red solid line is the total fit function. % (Color online)
 }
\label{fig:angles}
\end{center}
\end{figure}

The background-subtracted and efficiency-corrected distributions for $\cos\theta^*$, $\varphi_{\pm}$, and $\cos\theta_{\pm}$ 
are shown in Fig.~\ref{fig:angles}. We perform a simultaneous fit of 
Eqs.~\ref{phietaprodang},~\ref{phietaphihelicityang}, and~\ref{phietavarphiang} to the five angular distributions,
assuming there are no correlations between the variables. 
We  return to the issue of correlations when we discuss systematic uncertainties.
In the fit, the amplitudes are constrained to satisfy $|F_{00}|^2+4|F_{10}|^2+2|F_{11}|^2=1$,
since there are one $F_{00}$, four $F_{10}$ and two $F_{11}$ amplitude components.
The free parameters in the fit are $|{F_{00}|^2}$, $|{F_{10}|^2}$ and the overall normalization.
The value and error of $|{F_{11}|^2}$ is derived 
from the fit result and its full covariance matrix 
using $F_{11} = \frac{1}{2}(1-|{F_{00}}|^2-4|{F_{10}}|^2)$. 
The normalized amplitudes are: $\Myamplitudes$.

To determine the significance of the result and the systematic errors in the fitting procedure, 
we performed fits to multiple sets of events generated according to  Eq.~\ref{phietaang} (toy MC).
These studies allow us to assess biases that arise because of correlations.
We find the biases in the fitted ratios of squared moduli to be less than $0.002$, 
which are included in the systematic errors. 
Most of these biases are due to the imperfect MC efficiency corrections  that 
result from the coarse bin size of the efficiency table.
The statistical uncertainties are scaled using the RMS of the pull distributions from the toy MC study.
The fitter underestimates the statistical uncertainties by approximately 6\%.
Other sources of systematic error, such as some of those described below for the cross section,
have little dependence on angle, and thus are expected to be relatively small. We neglect them.

The measured value of $|{F_{00}}|^2$  deviates from 1.0, the value predicted by QCD 
models~\cite{stanrule}.  
From the toy MC studies, we determine the statistical probability for this deviation to 
be less than 1 in 3000 experiments, corresponding
to 3.4 standard deviations.  Including systematic 
uncertainties, the significance is 3.1 standard deviations.
This suggests that the production may not be dominated by single-photon annihilation as naively expected.

The  cross section, including radiative corrections, for \epem\to$\rho^+\rho^-$ is calculated from
\begin{equation}
\sigma=\frac{N}{\L  \times  \varepsilon \times (1+\delta)}\  ,
\end{equation}
where $N$ is the number of $\rho^+\rho^-$ signal events extracted from the combined data,
$\L$ is the integrated luminosity, and
$\varepsilon$ is the signal efficiency obtained from MC simulation that uses the
fully differential angular distribution derived from the results of the form factor fit. 
The $\rho^+\rho^-$ signal efficiency in the fiducial region, 
without radiative corrections, is estimated to be 15.0\%.
The correction for initial state radiation, $1+\delta$, 
is calculated according to Ref.~\cite{radiation} and has the value $0.775$.
Assuming  single-photon production, the radiatively corrected cross section near $\sqrt{s}=10.58\gev$
for $m_{\rho^{\pm}}~<~1.5 $ \gevcc and within $|\cos\theta^*|<$ 0.8, $|\cos\theta_{\pm}|\! <\! 0.85$ is
$\Resultrhoprhom$.
Using Eq.~\ref{phietaang}, we can scale the cross section from our acceptance to the full angular ranges, 
which gives $\ResultrhoprhomExtend$, where the systematic error includes $\pm1.7~\mathrm{fb}$ 
due to the effect of the uncertainties in the amplitudes on the extrapolation.

To study the possibility that the observed signal arises from $\Upsilon(4S)\to\rho^+\rho^-$ 
decay, we scale the off-resonance signal to the on-resonance luminosity, and
subtract it from the on-resonance signal. The resulting number of
events, $35 \pm 135$, is consistent with zero. The corresponding branching
fraction for $\Upsilon(4S)\to \rho^+ \rho^-$ is $(8.1 \pm 29.0) \times 10^{-7}$.
The systematic errors, which may be estimated from those given below for the cross section, are
negligible for this branching fraction measurement. 
Restricting possible results to the physical region ($\ge 0$), 
the Bayesian 90\% confidence level upper limit is $5.7\times 10^{-6}$.

The systematic uncertainty on the \epem\to$\rho^+\rho^-$ cross section, due to uncertainties
in the angular distribution fit, is estimated by varying the amplitude values.
The systematic uncertainty from the two-dimensional fit is estimated from the difference
in yield obtained by allowing the mean and width of the $\rho$ resonance mass to vary in the fit.
The systematic uncertainties due to $\pi^{\pm}$ identification,  tracking, and $\pi^0$ 
efficiency are estimated based on measurements from control data samples.
The possible background from related modes with extra particles is estimated
by using extrapolations from  four-particle mass sidebands.
The systematic uncertainties are summarized in Table~\ref{tab:syserror}.

\begin{table}[!htb]
\caption{Systematic uncertainties on the fiducial region
cross section of
$e^+e^- \rightarrow \rho^+\rho^-$.}
\begin{center}
\begin{tabular}{lrr}
\hline
Source                        & Systematic uncertainty \% \\
\hline\hline
Amplitude fit &  5.2 \\
Two-dimensional fit & 2.0 \\
Particle Identification & 2.3 \\
Tracking efficiency & 1.0 \\
$\pi^0$ efficiency & 6.0 \\
$\rho^+\rho^- + X$ feed-down & 4.9 \\
Luminosity &  2.0 \\
Radiative corrections & 1.0 \\
\hline
Total                                        &  10.0     \\
\hline\hline
\end{tabular}
\label{tab:syserror}
\end{center}
\end{table}

In summary, we have presented the first observation of the exclusive production of $\rho^+\rho^-$ in
$e^+e^-$ interactions near $\sqrt{s}=10.58$ GeV and measured the
relative amplitudes of the three helicity components.
Assuming production through single-photon annihilation, 
the cross section is measured to be $\sigma(\epem\!\! \to\! \rho^+ \rho^-)\! =$$\ResultrhoprhomExtend$.
The 90\% confidence level upper limit
on the branching fraction $\cal{B}$$ ( \Upsilon(4S) \to \rho^+ \rho^-)$ is
$5.7 \times 10^{-6}$. Our result for the  $|{F}_{00}|^2$ amplitude is 
inconsistent by $3.1$~standard deviations 
with the predictions of QCD models that assume single-photon production, however, 
indicating that other mechanisms such as TVPA with FSI may be important.

We are grateful for the excellent luminosity and machine conditions
provided by our \pep2\ colleagues, 
and for the substantial dedicated effort from
the computing organizations that support \babar.
We wish to thank S. Brodsky, A. Goldhaber, and L. Dixon for helpful discussions.
The collaborating institutions wish to thank 
SLAC for its support and kind hospitality. 
This work is supported by
DOE
and NSF (USA),
NSERC (Canada),
CEA and
CNRS-IN2P3
(France),
BMBF and DFG
(Germany),
INFN (Italy),
FOM (The Netherlands),
NFR (Norway),
MES (Russia),
MEC (Spain), and
STFC (United Kingdom). 
Individuals have received support from the
Marie Curie EIF (European Union) and
the A.~P.~Sloan Foundation.

\end{document}